\documentstyle[preprint,aps,epsf]{revtex}
\begin{document}
\draft

\title{
\vspace{-3.0cm}
\begin{flushright}
{\normalsize UTHEP-314}\\
\vspace{-0.3cm}
{\normalsize 1995 }\\
\end{flushright}
\vspace*{2.0cm}
{\large Domain-wall fermions with $U(1)$ dynamical gauge fields}
\vspace*{0.5cm}}

\author{ S. Aoki and K. Nagai}
\address{Institute of Physics,  University of Tsukuba,
         Tsukuba, Ibaraki 305, Japan}

\date{\today}
\maketitle

\begin{abstract}
We have carried out a numerical simulation
of a domain-wall model in $(2+1)$-dimensions,
in the presence of a dynamical gauge field only in an extra dimension,
corresponding to the weak coupling limit of a ( 2-dimensional ) physical
gauge coupling.
Using a quenched approximation we have investigated this model at
$\beta_{s} ( = 1 / g^{2}_{s} ) =$ 0.5 ( ``symmetric'' phase), 1.0, and 5.0
(``broken'' phase), where $g_s$ is the gauge coupling constant
of the extra dimension.
We have found that
there exists a critical value of a domain-wall mass $m_{0}^{c}$
which separates a region with a fermionic zero mode on the domain-wall
from the one without it,
in both symmetric and broken phases.
This result suggests that the domain-wall method
may work for the construction of lattice chiral gauge theories.
\end{abstract}

\narrowtext
\section{Introduction}
\label{sec:int}
Construction of chiral gauge theories is one of the long-standing problems
of lattice field theories.
Due to the fermion doubling problems, a naively discretized lattice fermion
field yields $2^d$ fermion particles, half of one chirality and half of the
other, so that the theory becomes non-chiral\cite{nielnino}.
Several lattice approaches have been proposed, but so far none of them have
been proven to work successfully.

Kaplan has proposed a new construction of lattice chiral gauge
theories via domain-wall models\cite{kaplan}.
Starting from a vector-like gauge theory in $2k+1$-dimensions
with a fermion mass term being a shape of a domain-wall in the
(extra) $2k+1$-th dimension,
he showed in the weak gauge coupling limit that
a massless chiral state arises as a zero mode
bound to the $2k$-dimensional domain-wall
while all the doublers have large masses of the lattice cut-off scale.
It has been also shown that the model works well for smooth back-ground gauge
fields\cite{aokihirose,jansen}.

Two simplified variants of the original Kaplan's domain-wall model
have been proposed: an ``overlap formula''\cite{overlap} and a
``waveguide model''\cite{waveg}. Gauge fields appeared in these variants
are $2k$-dimensional and are independent of the extra $2k+1$-th coordinate,
while those in the original model are $2k+1$-dimensional and depend on
the extra $2k+1$-th coordinate. These variants work successfully for
smooth back-ground gauge fields\cite{constchi,aokilevin}, as the original one
does. Non-perturbative investigations for these variants
seems easier than for the original model due to the simpler structure
of gauge fields.

However it has been reported\cite{waveg} that
the waveguide model in the weak gauge coupling limit
can not produce chiral zero modes needed
to construct chiral gauge theories.
In this limit, if gauge invariance were maintained,
pure gauge field configurations equivalent to the unity
by gauge transformation would dominate and gauge fields would become
smooth.
In the set-up of the waveguide model, however, $2k$-dimensional gauge
fields are non-zero only in the layers near domain-wall( waveguide ),
so that the gauge invariance is broken in the edge of the waveguide.
Therefore, even in the weak gauge coupling limit,
gauge fields are no more smooth and becomes very ``rough'',
due to the gauge degrees of freedom appeared to be dynamical in this edge.
As a result of the rough gauge dynamics,
a new chiral zero mode with the opposite chirality to the original zero mode
on the domain-wall appears in the edge, so that
the fermionic spectrum inside the waveguide becomes vector-like.
It has been claimed\cite{waveg} that this ''rough gauge'' problem also
exists in the overlap formula since the gauge invariance is broken
by the boundary condition at the infinity of the extra
dimension\cite{aokilevin,shamir}. Furthermore an equivalence between the
wave-guide model and the overlap formula has been pointed out for
the special case\cite{gswgol}. Although the claimed equivalence
has been challenged in ref.\cite{overwave}, it is still crucial
for the success of the overlap formula to solve
the ''rough gauge'' problem and to show the existence of a chiral zero mode
in the weak gauge coupling limit.

How about original Kaplan's model ?
In this model there are two inverse gauge coupling $\beta=1/g^2$ and
$\beta_s=1/g_s^2$, where $g$ is the coupling constant in (physical)
$2k$-dimensions and $g_s$ is the one in the (extra) $2k+1$-th dimension.
Very little are known about this model except $\beta_s=0$
case\cite{aoitnioshi,kornip} where the spectrum seems vector-like.
In the weak coupling limit, corresponding to the $g\rightarrow 0$ limit in this
model, all gauge fields in the physical dimensions can be gauged away, while
the gauge field in the extra dimension is still dynamical and its dynamics is
controlled by $\beta_s$. Instead of the gauge degrees of freedom
in the edge of the wave-guide, $2k+1$-th component of gauge fields represent
roughness of $2k$ dimensional gauge fields.
An important question is whether the chiral zero mode on the domain-wall
survives in the presence of this rough dynamics.
The dynamics of the gauge field in this limit is equivalent to
$2k$ dimensional scalar model with $L_s$ independent copies where $L_s$ is the
number of sites in the extra dimension. In general at large $\beta_s$ such a
system is in a ``broken phase'' where some global symmetry is spontaneously
broken, while at small $\beta_s$ the system is in a ``symmetric'' phase.
Therefore there exists a critical point $\beta_s^c$, and it is likely
that the phase transition at $\beta_s=\beta_s^c$ is continuous(second or higher
order).
The ''gauge field'' becomes rougher and rougher at smaller $\beta_s$.
Indeed we know that the zero modes disappears at $\beta_s=0$\cite{aoitnioshi},
while the zero mode exists at $\beta_s=\infty$ ( free case ).
So far we do not know the fate of the chiral zero mode in the intermediate
range of the coupling $\beta_s$. There are the following 3 possibilities:
(a) The chiral zero mode always exists except $\beta_s=0$. In this case
we may likely construct a lattice chiral gauge theory
in both broken ($\beta_s > \beta_s^c$) and symmetric($\beta_s < \beta_s^c$),
and the continuum limits may be taken at $\beta_s = \beta_s^c$.
This is the beset case for the domain-wall model.
(b) The chiral zero mode exists only in the broken
phase($\beta_s > \beta_s^c$). In this case we may construct a lattice chiral
gauge theory only in the broken phase via the domain-wall method.
This is unsatisfactory, since the chiral gauge theory in the symmetric phase,
which is an important theoretical basis for various models,
can not be described via the domain-wall method.
(c) No chiral zero mode survives except $\beta_s=\infty$. The original
Kaplan's model can not describe lattice chiral gauge theories at all.

It is very important to determine which possibility is indeed realized
in the domain-wall model.
Therefore, in this paper ,
in order to know the fate of the chiral zero mode,
we have carried out a numerical simulation of a domain-wall model
in $(2+1)$-dimension
with a quenched $U(1)$ gauge field in the $\beta=\infty$ limit.
Strictly speaking, there is no order parameter in a 2 dimensional
$U(1)$ model( XY model ). On a large but finite lattice, however, a behavior
of the 2 dimensional model is similar to the one of a 4 dimensional
scalar model. Thus, we hopefully think that
useful informations about the fate of the zero mode
can be obtained from such a toy model in (2+1)-dimensions.
In Sec.2 ,
we have defined our domain-wall model
with dynamical gauge fields.
We have calculated a fermion propagator
by using a kind of mean-field approximation,
to show that there is a critical value of the domain-wall mass
parameter above which the zero mode exist.
The value of the critical mass may depend on $\beta_s$,
which controls the dynamics of the gauge field.
In Sec.3 ,
we have calculated the fermion spectrum numerically
using quenched approximation at $\beta_s = 0.5,1.0.5.0$
and at various values of domain-wall masses.
We have found that at any value of three $\beta_s$
there always exists the range of domain-wall mass parameter
in which the chiral zero mode survives on the domain-wall.
Our conclusion and some discussions are given in Sec. 4.
%
%
%
%
\section{domain-wall model}
\subsection{Definition of the model}
We consider a vector gauge theory in $d=(2k+1)$-dimension
with a domain-wall mass term, which has a shape of a step function
in the coordinate of an extra dimension.
This  domain-wall model is originally proposed by Kaplan\cite{kaplan},
and a fermionic part of the action
is reformulated by Narayanan-Neuberger\cite{naraneu},
in terms of a $2k$-dimensional theory.
The model is defined by the action
\begin{equation}
S = S_{G} + S_{F} ,
\end{equation}
where $S_{G}$ is the action of a dynamical gauge field , $S_{F}$ is the
fermionic action.
$S_{G}$ is given by
\begin{eqnarray}
S_{G} &=& \beta \sum_{n,\mu > \nu} \sum_{s} {\rm Re Tr} \left[
U_{\mu \nu}(n,s) \right] \nonumber \\
&+& \beta_{s} \sum_{n,\mu} \sum_{s} {\rm Re Tr} \left[
U_{\mu d}(n,s) \right] ,
\end{eqnarray}
where $\mu ,  \nu$ run from $1$ to $2 k$ ,
$n$ is a $2 k$-dimensional lattice point ,
and $s$ is a coordinate of an extra dimension.
$U_{\mu \nu}(n,s)$ is a $2 k$-dimensional plaquette
and $U_{\mu d}(n,s)$ is a plaquette
containing two link variables in the extra direction.
$\beta$ is the inverse gauge coupling for the plaquette $U_{\mu \nu}$
and $\beta_{s}$ is the one for the plaquette $U_{\mu d}$ .
In general , $\beta \neq \beta_{s}$ .
The fermion action $S_{F}$ on the Euclidean lattice,
in terms of the $2 k$-dimensional notation,
is given by
\widetext
\begin{eqnarray}
 S_{F}& = & \frac{1}{2} \sum_{n \mu} \sum_s
        \bar{\psi}_s(n) \gamma_\mu \left[ U_{s,\mu}(n) \psi_s(n + \mu)
        - U_{s,\mu}^{\dag}(n - \mu) \psi_s(n - \mu) \right]  \nonumber \\
&+& \sum_n \sum_{s,t} \bar{\psi}_s(n) \left[ M_0 P_R
        + M_0^{\dag} P_L \right] \psi_t(n)  \nonumber \\
&+& \frac{1}{2} \sum_{n \mu} \sum_s
        \bar{\psi}_s(n) \left[ U_{s,\mu}(n) \psi_s(n + \mu)
        + U_{s,\mu}^{\dag}(n - \mu) \psi_s(n - \mu) -2 \psi_s(n) \right]
\label{eqn:fermion}
\end{eqnarray}
\narrowtext
where $s , t$ are an extra coordinates ,
$P_{R/L} = \frac{1}{2} (1 \pm \gamma_{2k+1})$ ,
\begin{eqnarray}
&& ( M_0 )_{s,t} = U_{s,d}(n) \delta_{s + 1 , t} - a(s) \delta_{s,t} \\
&& ( M_0^{\dag} )_{s,t} = U_{s - 1 , d}^{\dag}(n) \delta_{s - 1 , t} - a(s)
\delta_{s,t} .
\end{eqnarray}
Here $U_{s,\mu}(n) , U_{s,d}(n)$ ($d=2k+1$) are link variables
connecting a site $(n,s)$ to $(n+\mu,s)$ or $(n,s+1)$, respectively,
Because of a periodic boundary condition in the extra dimension ,
$s , t$ run from $-L_{s}$ to $L_{s} - 1$ ,
and $a(s)$ is given by
\begin{eqnarray}
  a(s) & = & 1 - m_0 \, {\rm{sign}}
\left[( s + \frac{1}{2} ) \, {\rm{sign}}( L_{s} - s - \frac{1}{2} ) \right]
\nonumber \\
 & = & \left\{
\begin{array}{cc}
1 - m_0  &  ( - \frac{1}{2} < s < L_{S} - \frac{1}{2} ) \\
1 + m_0  &  ( - L_{s} - \frac{1}{2} < s < - \frac{1}{2} ) \, ,
\end{array}
\right.
\end{eqnarray}
where $m_{0}$ is the height of the domain-wall mass.
It is easy to check
that the above fermionic action is identical to the one
in $(2k+1)$-dimensions, proposed by Kaplan\cite{kaplan,naraneu}.

In weak coupling limit of both $\beta$ and $\beta_{s}$ ,
this model becomes free theory
and can be easily analyzed.
In free theory at $0 < m_{0} < 1$,
it has been shown that a desired chiral zero mode
appears on a domain-wall( $s=0$ plane ) without unwanted doublers.
Due to the periodic boundary condition in the extra dimension,
however , a zero mode of the opposite chirality to the one on the domain-wall
appears on the anti-domain-wall , $s = L_{s} - 1$ .
Overlap between two zero modes decreases exponentially at large $L_s$.
A free fermion propagator is easily calculated and
an effective action of a $(2+1)$-dimensional model
including the gauge anomaly and the Chern-Simons term can be obtained for
smooth background gauge fields\cite{aokihirose}.

Domain-wall models, however, have not been investigated yet
{\it non-perturbatively}.
Main question is whether the chiral zero mode survives
in the presence of rough gauge fields mentioned in the introduction.
To answer this question we will analyze
the fate of the chiral zero mode
in the weak coupling limit for $\beta$.
In this limit,
the gauge field action $S_{G}$ is reduced to
\begin{equation}
S_{G} = \beta_{s} \sum_{s} \sum_{n} {\rm Re Tr} \left[ V(n,s) V^{\dag}(n+\mu,s)
\right] ,
\label{eqn:gauge}
\end{equation}
where the link variable $U_{s,d}(n)$ in the extra direction
is regarded as a site variable $V(n,s)( = U_{s,d}(n))$.
This action is identical to the one of a $(d-1)$-dimensional spin model
and $s$ is regarded as an independent flavor.
The action eq.(\ref{eqn:gauge}) is invariant under
\begin{equation}
V(n,s) \longrightarrow g(s) V(n,s) g^{\dag}(s+1)\quad, \qquad
(g(s) \in G) ,
\label{eqn:symmetry}
\end{equation}
where $G$ is the gauge group of the original model.
Therefore the total symmetry of the model
is $G^{2 L_{s}}$, where $2L_s$, the size of the extra
dimension, is regarded as the number of independent flavors.
We use this (reduced) model for our numerical investigation.
%
%
%
\subsection{Mean field approximation for fermion propagators}
When the dynamical gauge fields are added even on the extra dimension only,
it is difficult to calculate the fermion propagator analytically.
Instead of calculating the fermion propagator {\it exactly} ,
we use a mean-field approximation
to see an effect of the dynamical gauge field qualitatively.
The mean-field approximation we adopt is
that the link variables are replaced as
\begin{equation}
V(n,s) \longrightarrow z ,
\end{equation}
where $z$ is a $(n,s)$-independent constant.
{}From eq.(\ref{eqn:fermion})
the fermion action in a $(d-1)$-dimensional momentum space becomes
\widetext
\begin{equation}
S_{F} \rightarrow
\sum_{s,t,p} \bar{\psi}_{s}(-p)
\left(\sum_{\mu} i \gamma_{\mu} \sin (p_{\mu}) \delta_{s,t}
+ \left[ M(z) P_{R} + M^{\dag}(z) P_{L} \right]_{s,t} \right)
\psi_{t}(p),
\end{equation}
\begin{equation}
(M(z))_{s,t} = (M_{0}(z))_{s,t} + \frac{\nabla(p)}{2} \delta_{s,t} ,
\qquad
(M^{\dag}(z))_{s,t} = (M^{\dag}_{0}(z))_{s,t} + \frac{\nabla(p)}{2}
\delta_{s,t} ,
\end{equation}
where
$ \nabla(p) \equiv \sum_{\mu=1}^{d-1} 2 ( \cos p_{\mu} - 1 ).$
Following Ref.\cite{aokihirose} it is easy to obtain
a mean field fermion propagator
on a finite lattice with the periodic boundary condition:
\begin{eqnarray}
 G(p)_{s,t}
&=&
\left[ i \sum_\mu \gamma_\mu \bar{p}_\mu + M P_R + M^{\dag}
P_L \right]_{s,t}^{-1}
\nonumber \\
&=&
\left[ \left\{ \left(
        - i \sum_\mu \gamma_\mu \bar{p}_\mu + M \right)
G_L(p)_{s,t} \right\} P_L
        \right.
+ \left.  \left\{ \left(
         - i \sum_\mu \gamma_\mu \bar{p}_\mu + M^{\dag} \right)
G_R(p)_{s,t} \right\} P_R \right] ,
\label{eqn:propagator}
\end{eqnarray}
\narrowtext
\begin{equation}
 G_L(p) = \frac{\displaystyle 1}{\displaystyle {\bar{p}^2 + M^{\dag} M}}
\quad , \quad
 G_R(p) = \frac{\displaystyle 1}{\displaystyle {\bar{p}^2 + M M^{\dag}}} \quad
,
\label{eqn:fomal}
\end{equation}
with $\bar{p}_{\mu} = \sin (p_{\mu})$.
For large $L_{s}$ where we neglect terms of $O(e^{-c L_{s}})$ with $c > 0$,
$G_{L}$ and $G_{R}$ are given by
\widetext
\begin{equation}
 \left[ G_L(p) \right]_{s,t} = \left\{
\begin{array}{ll}B e^{-\alpha_{+} |s - t|}
        + \left( A_L - B \right) e^{- \alpha_{+}(s + t)}
        + \left( A_R - B \right) e^{- \alpha_{+}(2L_{s} - s - t)} ,
 & (s , t \geq 0)  \\
A_L e^{- \alpha_{+} s + \alpha_{-} t}
        + A_R e^{- \alpha_{+}(L_{s} - s) - \alpha_{-}(L_{s} + t)} ,
 & (s \geq 0 , t \leq 0) \\
A_L e^{ \alpha_{-} s - \alpha_{+} t}
        + A_R e^{- \alpha_{-}(L_{s} + s ) - \alpha_{+}(L_{s} - t)} ,
 & (s \leq 0 , t \geq 0) \\
C e^{-\alpha_{-} |s - t|}
        + \left( A_L - C \right) e^{ \alpha_{-}(s + t)}
        + \left( A_R - C \right) e^{- \alpha_{-}(2L_{s} + s + t)},
 & (s , t \leq 0)
\end{array}
\right.
\label{eqn:GL}
\end{equation}
\begin{equation}
 \left[ G_R(p) \right]_{s,t} = \left\{
\begin{array}{ll}
B e^{-\alpha_{+} |s - t|}
        + \left( A_R - B \right) e^{- \alpha_{+}(s + t + 2)}
        + \left( A_L - B \right) e^{- \alpha_{+}(2L_{s} - s - t -2)} ,
 & (s , t \geq -1)  \\
A_R e^{- \alpha_{+}(s + 1) + \alpha_{-} (t + 1)}
        + A_L e^{- \alpha_{+}(L_{s} - s -1) - \alpha_{-}(L_{s} + t +1)} ,
 & (s \geq -1 , t \leq -1) \\
A_R e^{ \alpha_{-} (s + 1) - \alpha_{+} (t + 1)}
        + A_L e^{- \alpha_{-}(L_{s} + s + 1) - \alpha_{+}(L_{s} - t -1)} ,
 & (s \leq -1 , t \geq -1) \\
C e^{-\alpha_{-} |s - t|}
        + \left( A_R - C \right) e^{ \alpha_{-}(s + t +2)}
        + \left( A_L - C \right) e^{- \alpha_{-}(2L_{s} + s + t +2)} ,
 & (s , t \leq -1)
\end{array}
\right.
\label{eqn:GR}
\end{equation}
\narrowtext
where
\begin{eqnarray}
&& a_{\pm} = z ( 1 - \frac{\nabla(p)}{2} \mp m_{0} ) = z b_{\pm} , \\
&& \alpha_{\pm} = {\rm{arccosh}}
 \left[ \frac{\bar{p}^{2} + z^{2} + b_{\pm}^2}{2 z b_{\pm}} \right] , \\
&& A_L = \frac{1}{a_{+} e^{\alpha_{+}} - a_{-} e^{- \alpha_{-}}}, \,
   A_R = \frac{1}{a_{-} e^{\alpha_{-}} - a_{+} e^{- \alpha_{+}}}, \\
&& B = \frac{1}{2 a_{+} \sinh \alpha_{+}} \quad , \quad
   C = \frac{1}{2 a_{-} \sinh \alpha_{-}} .
\end{eqnarray}

Behaviors of $A_{R} , B$ and $C$ as $p\rightarrow 0$
are similar to the ones in free theory:
They have no singularity for all $z$
A behavior of $A_{L}$ is, however, different:
As $p \rightarrow 0$ $A_{L}$ behaves as
\begin{eqnarray}
A_{L} & \rightarrow &
\frac{\displaystyle 1}{\displaystyle [(1 - m_{0})^{2}] + O(p^{2})}, \quad
(0 < m_{0} < 1 - z), \\
& \rightarrow &
\frac{\displaystyle 4m_{0}^{2}-[(z^{2}-1)-m_{0}^{2}]^{2}}
{\displaystyle 4m_{0}z^{2}p^{2}}, \, \, (1 - z < m_{0} < 1) .
\end{eqnarray}
A critical value of the domain-wall mass
that separates a region with a zero mode
and a region without zero modes
is $m_{0}^c = 1 - z$.
Since $A_{L}$ term dominates for $1 - z  < m_{0} < 1$
in the $G_{L}$ (eq.(\ref{eqn:GL}) ) and $G_{R}$ ( eq.(\ref{eqn:GR})),
a right-handed zero mode appears in the $s=0$ plane ,
and a left-handed zero mode in the $s=L_{s}-1$ plane.
For $0 < m_{0} < 1 - z$
the right- and left-handed fermions are massive in all $s$ planes.
Since the terms of $A_{L} , A_{R} , B$ and $C$ are almost same value
in this region of $m_{0}$,
a translational invariant term dominates
in $G_{L}$ and $G_{R}$, so that
the spectrum becomes vector-like.

If $z \rightarrow 1$ , the model becomes free theory.
The propagator obtained in this section agrees with the one
obtained in Ref.\cite{aokihirose}.
In the opposite limit that $z \rightarrow 0$ ,
since there is no hopping term to the neighboring layers,
this model becomes the one analyzed in Ref.\cite{aoitnioshi}
in the case of the strong coupling limit $\beta_{s} = 0$ ,
and in Ref.\cite{kornip} ,
in the case that $z$ is identified to the vacuum expectation value of the link
variables.
This consideration suggests that
the region where the zero modes exist
become smaller and smaller as $z$ $(1-z < m_{0} < 1)$ approaches zero.
What corresponds to $z$ ?
Boundary conditions $z$ satisfies are
$z=1$ at $\beta_s=\infty$ and $z=0$ at $\beta_s=0$.
The most naive candidate\cite{kornip} is
\begin{equation}
z= \langle V(n,s) \rangle .
\label{order}
\end{equation}
But this is not invariant under the symmetry (\ref{eqn:symmetry}).
The other choice invariant under (\ref{eqn:symmetry}) is
\begin{equation}
z^2 = \langle {\rm Tr Re} \{V(n,s) V^\dagger(n+\mu,s)\} \rangle .
\label{energy}
\end{equation}
If eq. (\ref{order}) is true, zero modes disappears in the symmetric phase,
where $\langle V(n,s) \rangle = 0$, while, for the case of eq. (\ref{energy}),
the zero modes always exist in both phases, since
$\langle {\rm Tr Re}\{ V(n,s) V^\dagger(n+\mu,s)\} \rangle$
is insensitive to which phase
we are in.
%
%
%
%
\section{numerical study of (2+1)-dimensional U(1) model}
\subsection{Method of numerical calculations}
In this section
we numerically study the domain-wall model in $(2+1)$-dimension
with a $U(1)$ dynamical gauge field in the extra dimension.
As seen from eq.(\ref{eqn:gauge}) ,
the gauge field action can be identified with a $2$-dimensional $U(1)$ spin
model (with $2 L_{s}$ copies).
In $(2+1)$-dimension,
$\gamma$-matrices are Pauli-matrices ,
$\sigma_{1} \, , \, \sigma_{2} \, , \, \sigma_{3}$.

Our numerical simulation has been carried out by the quenched approximation.
Configurations of $U(1)$ dynamical gauge field are generated
and fermion propagators are calculated on the configurations.
The obtained fermion propagators are gauge non-invariant in general
under the symmetry (\ref{eqn:symmetry}).
The fermion propagator $G(p)_{s,t}$ becomes ``invariant''
if and only if $s=t$.
Thus, we take the $s-s$ layer as propagating plane($\approx$
``physical space''),
and investigate the behavior of the fermion propagator in this layer.

To study the fermion spectrum,
we assume a form of eq.(\ref{eqn:propagator}) for
the fermion propagator, from which we extract $G_{L}$ and $G_{R}$.
We then obtain corresponding fermion masses
from $G_{L}^{-1}(p)$ and $G_{R}^{-1}(p)$
by fitting them linearly in $\bar{p}^{2}$, since,
from eq.(\ref{eqn:fomal}) :
\begin{eqnarray}
G_{L}^{-1} = \bar{p}^{2} + M^{\dag} M \rightarrow m_{f}^{2}
\quad , \quad (p \rightarrow 0) ,
\label{eqn:Lanalysis} \\
G_{R}^{-1} = \bar{p}^{2} + M M^{\dag} \rightarrow m_{f}^{2}
\quad , \quad (p \rightarrow 0) .
\label{eqn:Ranalysis}
\end{eqnarray}
We take the following setup for 2-dimensional momenta.
A periodic boundary condition is taken for the 1st direction
and the momentum in this direction is fixed on $p_{1} = 0$.
An anti-periodic boundary condition is taken for
the 2nd-direction
and the momentum in this direction is variable
such as $p_{2} = (2n+1) \pi / L \, , \, n = - L / 2 ,..., L / 2 - 1$.)

If $m_{f}^{2} = 0$ ,
we conclude that there is a zero mode, and if $m_{f}^{2} \neq 0$ ,
there is not.
%
%
\subsection{Simulation parameters}
Our simulation is performed in the quenched approximation on
$L^{2}\times 2 L_{s}$ lattices with $L = 16 , 24 , 32$ and $L_{s} = 16$.
The coordinate $s$ in the extra dimension runs $-16 < s < 15$.
Gauge configurations are generated by the 5-hit Metropolis algorithm
at $\beta_{s} =$ 0.5, 1.0, 5.0. For the thermalization first 1000 sweeps
are discarded.

The fermion propagators are calculated by the conjugate gradient method
on 50 configurations separated by at least 20 sweeps, except at $\beta_s=$5.0
on a $32^2\times 32$ lattice where the number of configurations are 11.
We take the domain-wall mass
$m_{0} =$ 0.7, 0.8, 0.9, 0.99 at $\beta_{s} =$ 0.5,
$m_{0} =$ 0.3, 0.4, 0.5, 0.6, 0.9 at $\beta_{s} =$ 1.0,
and $m_{0} =$ 0.1, 0.2, 0.3 at $\beta_{s} =$ 5.0 .
The boundary conditions in $1$st- and $3$rd(extra)-directions
are periodic and the one in $2$nd-direction is anti-periodic.
Wilson parameter $r$ has been set to $r = 1$.
The fermion propagators have been investigated at $s =$ 0, 8, 15.
These $s$ are the layers where we put sources.
The layer at $s=0$ is the domain-wall ,
at $s=15$ , the anti-domain-wall ,
and at $s=8$ , neither.
Errors are all estimated with the jack-knife method.

\subsection{Quenched phase structure}

As explained before
the gauge field action of our model is identical to
that of the $U(1)$ spin system in $2$-dimensions.
Therefore, there is a Kosterlitz-Thouless phase transition
and this system does not have an order parameter on the infinite lattice.
On the finite volume, however,
we take a vacuum expectation value of link variables as an order parameter
using rotation technique:
\begin{equation}
v = < |\frac{1}{L^{2}} \sum_{n} V(n,s)| >_{s} ,
\label{eqn:order}
\end{equation}
where $L$ is the lattice size of the $1,2$-dimension.

\begin{figure}
\centerline{\epsfxsize=12cm \epsfbox{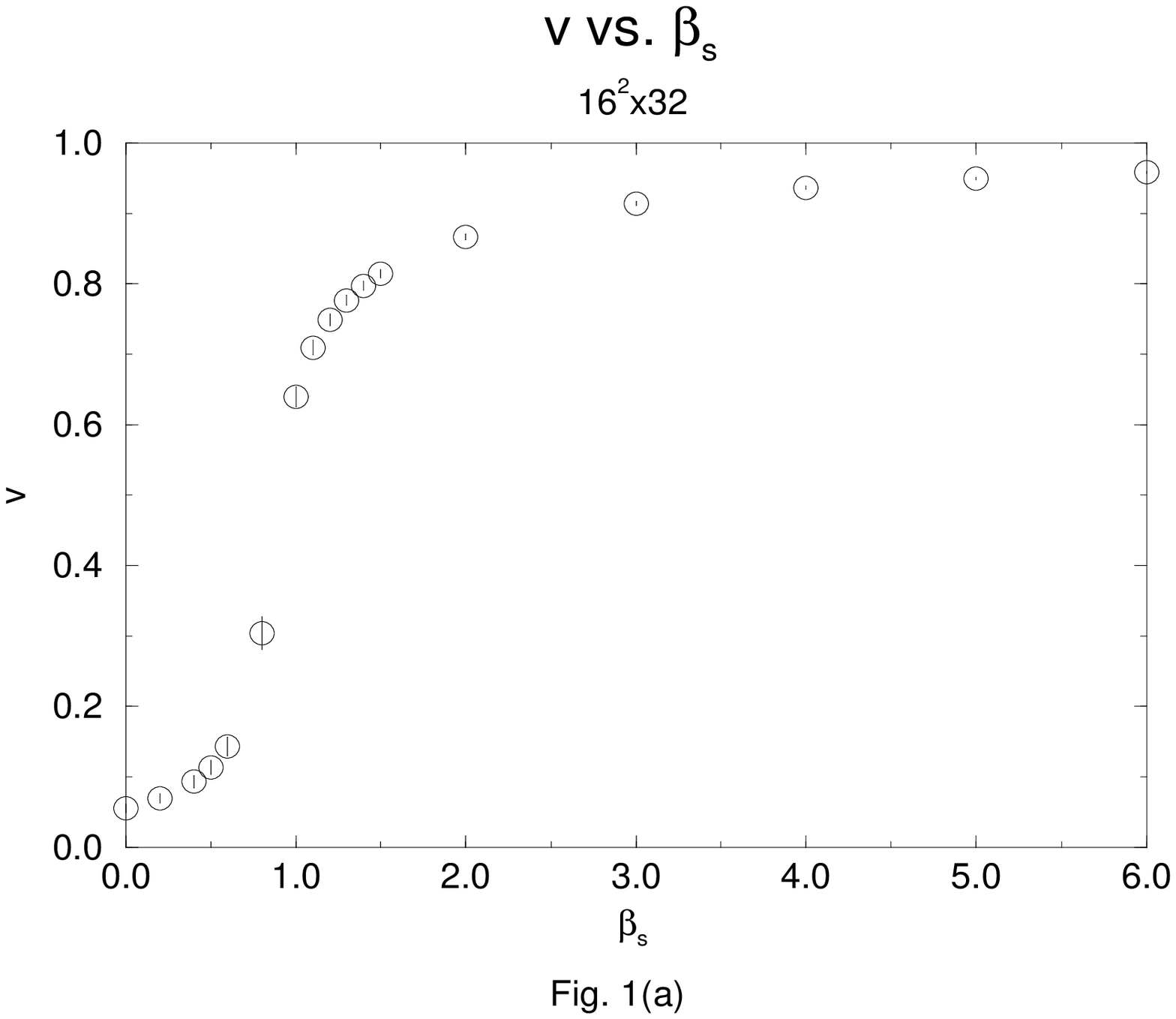}}
\centerline{\epsfxsize=12cm \epsfbox{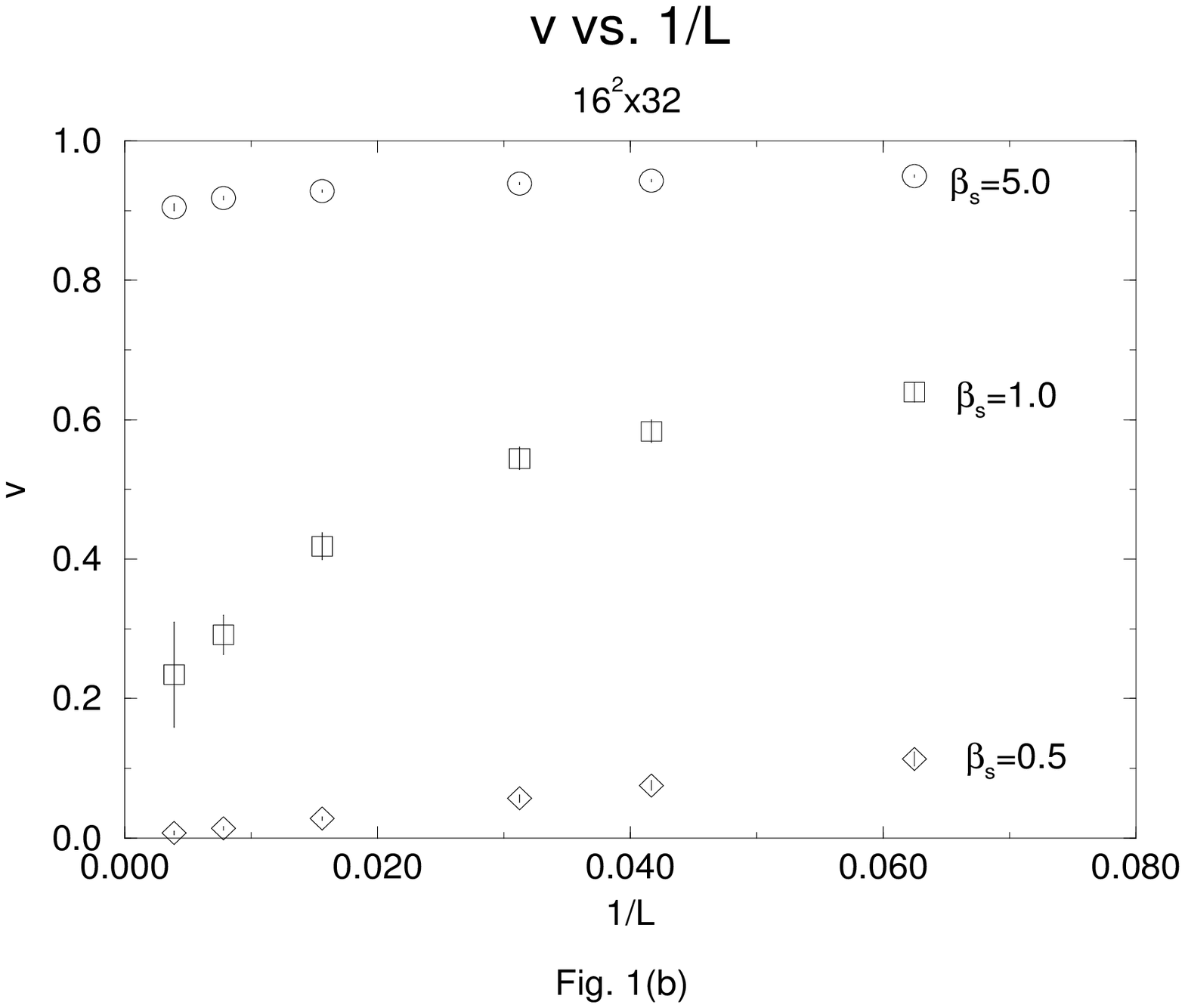}}
\caption{
(a) Vacuum expectation value of link variables $v$
on a $16^2\times 32$ lattice as a function of $\beta_s$.
(b) A volume dependence of the vacuum expectation values of link variables $v$
 as a function of $1/L$. }
\label{vev}
\end{figure}

The defined vacuum expectation value $v$ above is zero
in the Kosterlitz-Thouless phase
but $v > 0$ in the spin-wave phase on the finite lattice.
(Increasing the lattice size, however , decreasing the value of $v$.
In the infinite lattice size, the value of $v$ is zero for all gauge coupling.)
Since we are interested in the dynamics of $4$-dimensional theories,
where the phase transition separates a symmetric phase from a broken phase ,
we have used
this $2$-dimensional system on large but finite volume
as a toy model of $4$-dimensional real world.
Therefore, in this letter,
we refer to the Kosterlitz-Thouless phase as the symmetric phase,
and to the spin-wave phase as the broken phase.
Fig. \ref{vev}(a) shows that,
on a $16^2\times 32$ lattice, $v$ behaves as if it was an order parameter.
{}From Fig. \ref{vev}(b)
we consider that the system is in the symmetric phase at $\beta_{s} = 0.5$,
while in the broken phase at $\beta_{s} =$ 1.0, 5.0 .
%
%
\subsection{Fermion spectrum in the broken phase}
At $\beta_{s} =$ 1.0 and 5.0,
the system is in broken phase.
Here we mainly discuss the result at $\beta_{s} = 1.0$ in detail.

\begin{figure}
\centerline{\epsfxsize=12cm \epsfbox{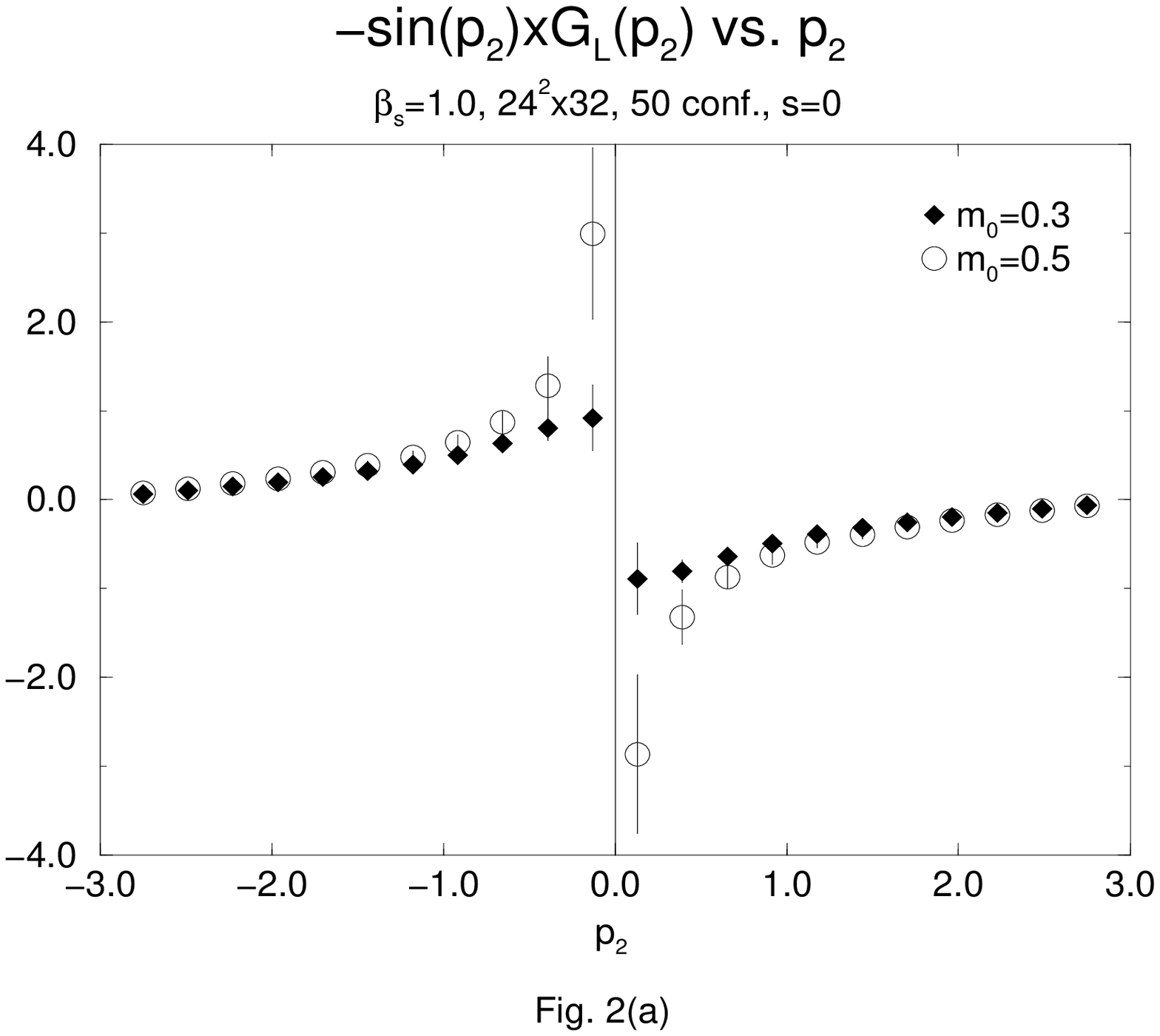}}
\centerline{\epsfxsize=12cm \epsfbox{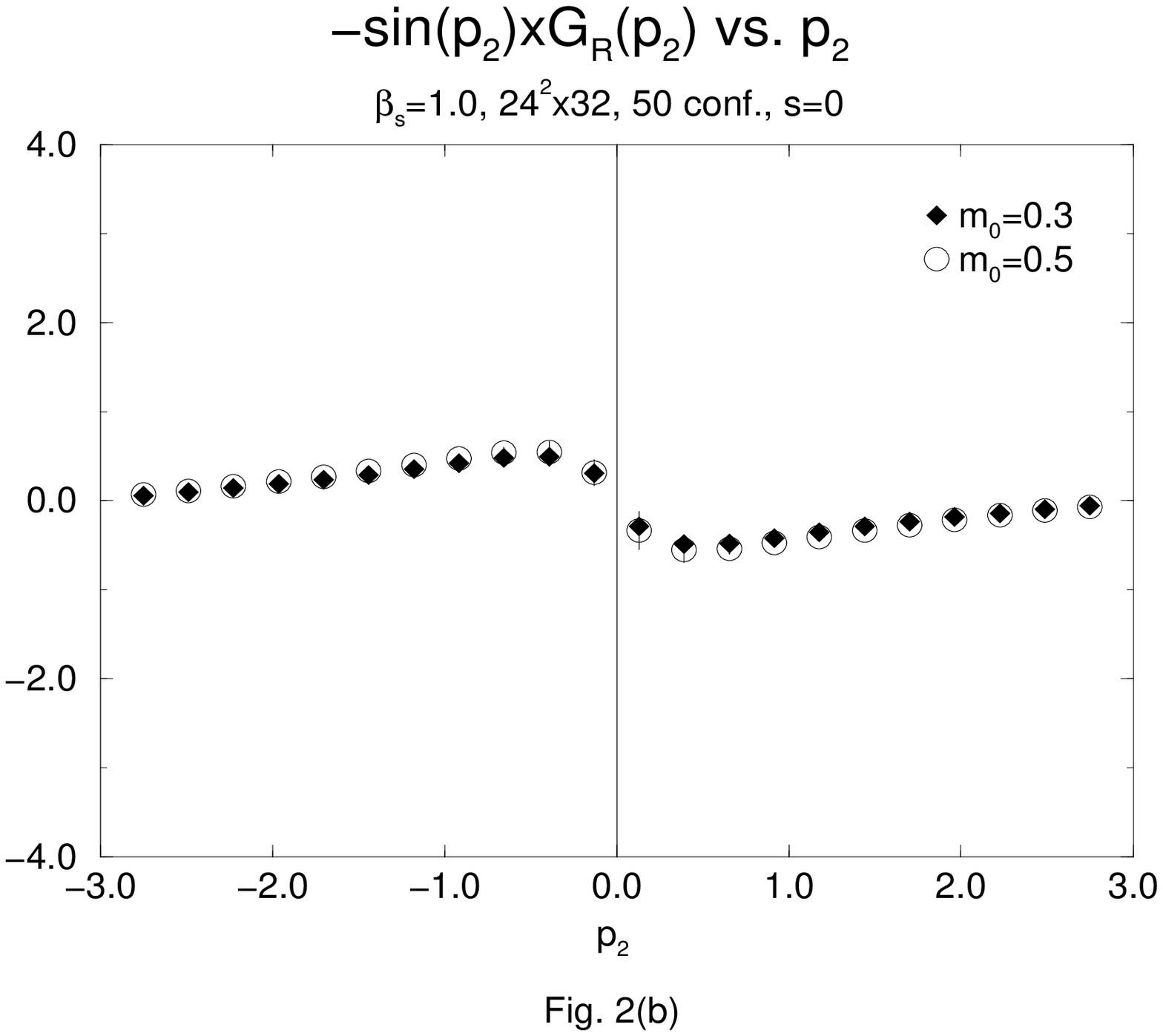}}
\caption{
$-\sin(p_2)\cdot [G_L]_{0,0}$
and $-\sin(p_2)\cdot [G_R]_{0,0}$
in the fermion propagator as a function of $p_2$
with $p_1=0$
at $\beta_{s}=1.0$ on a $24^2 \times 32$ lattice,
for $m_0$=0.5(open circles) and 0.3(solid diamonds).}
\label{prop}
\end{figure}

\begin{figure}
\centerline{\epsfxsize=12cm \epsfbox{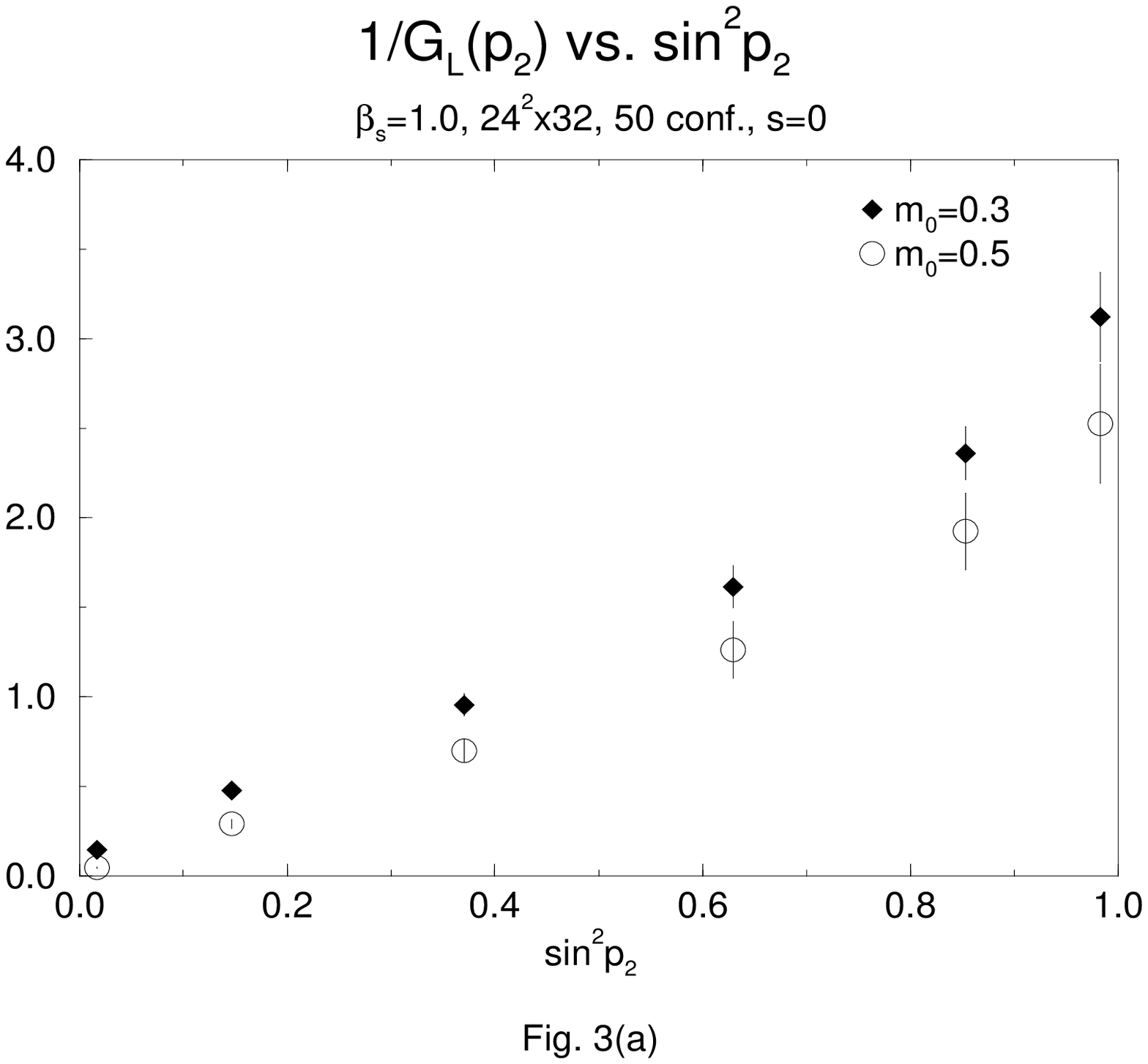}}
\centerline{\epsfxsize=12cm \epsfbox{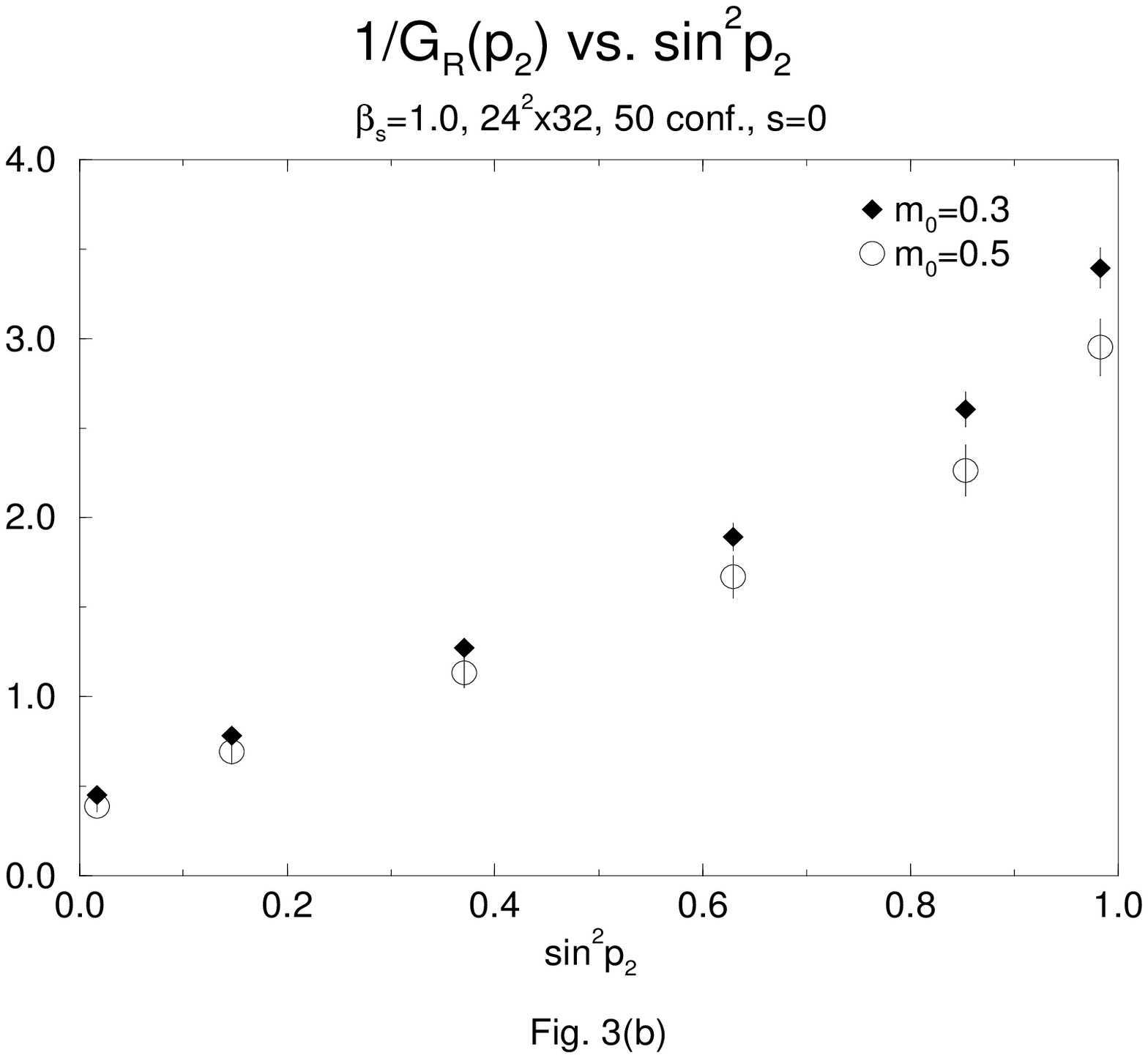}}
\caption{
$[G_L]_{0,0}^{-1}$
and $[G_{R}]_{0,0}^{-1}$ as a function of $\sin^{2}(p_{2})$
with $p_1=0$
at $\beta_{s}=1.0$ on a $24^2 \times 32$ lattice,
for $m_0$=0.5(open circles) and 0.3(solid diamonds).}
\label{invprop}
\end{figure}

\begin{figure}
\centerline{\epsfxsize=12cm \epsfbox{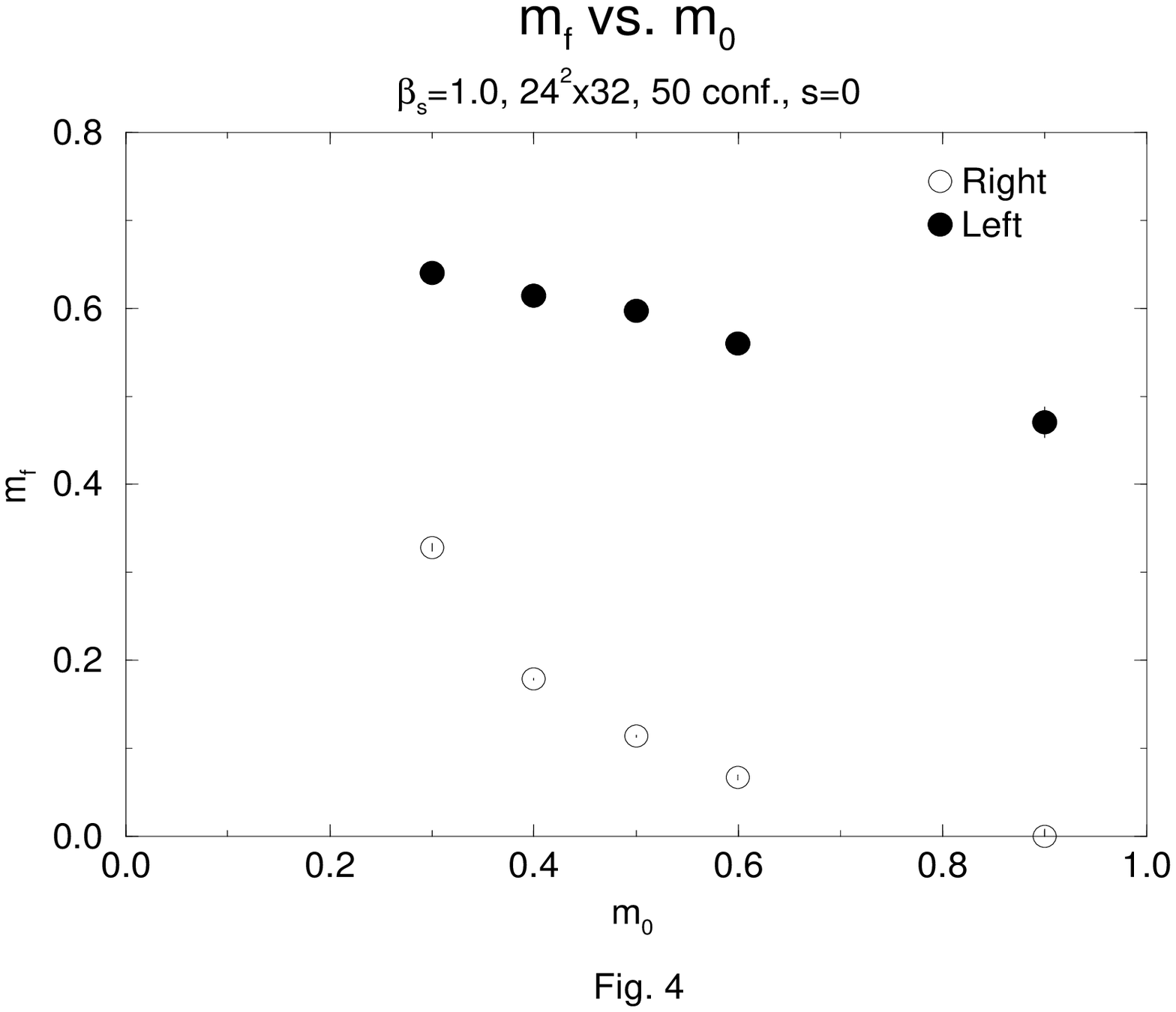}}
\caption{
$m_{f}$ vs. $m_{0}$
at $\beta_{s} = 1.0$ on a $24^{2} \times 32$ lattice,
in the case of putting a source on the domain-wall $s=0$,
for the right-handed fermion(open circles) and the left-handed
fermion(solid circles).}
\label{mfs0}
\end{figure}

We first consider the fermion spectrum on the layer at $s=0$.
Fig. \ref{prop} is a plot of the term corresponding to
$- \sin (p_{2})\cdot G_{L}$ and $- \sin (p_{2})\cdot G_{R}$
as a function of $p_{2}$ at $m_0$ =0.3 and 0.5.
( Note we always set $p_1=0$.)
This figure shows that, as $p_2$ goes to zero,
$G_{L}$ seems to diverge at $m_{0}$ =0.5 but stay finite at $m_{0}$ = 0.3,
while $G_{R}$ stays finite at both $m_0$.
Next let us show Fig. \ref{invprop}, which
is a plot of the $G_{L}^{-1}$ and $G_{R}^{-1}$ as a function of
$\bar{p}_2^{2} \equiv \sin^{2} (p_{2})$ at $m_0$ =0.3 and 0.5.
In the limit $p_{2} \rightarrow 0$,
$G_{R}^{-1}$ remains non-zero at both $m_{0}$,
while $G_{L}^{-1}$ vanishes at $m_{0} = 0.5$.
We obtain the value of $m_f^2$, which can be regarded as the mass
square in $2$-dimensional world,
by the linear fit in $\bar{p}_2^{2}$,
and plot $m_f$ as a function of $m_{0}$ in Fig. \ref{mfs0}.
The mass of right-handed fermion, obtained from $G_{L}^{-1}$,
becomes very small ( less than 0.1) at $m_{0}$ larger than $0.5$, so
we conclude that this critical value is $m_{0}^{c} \sim 0.5$.
Whenever the domain-wall mass is larger than this value ,
this model produces the right-handed chiral zero mode on the domain-wall
at $s=0$.

\begin{figure}
\centerline{\epsfxsize=12cm \epsfbox{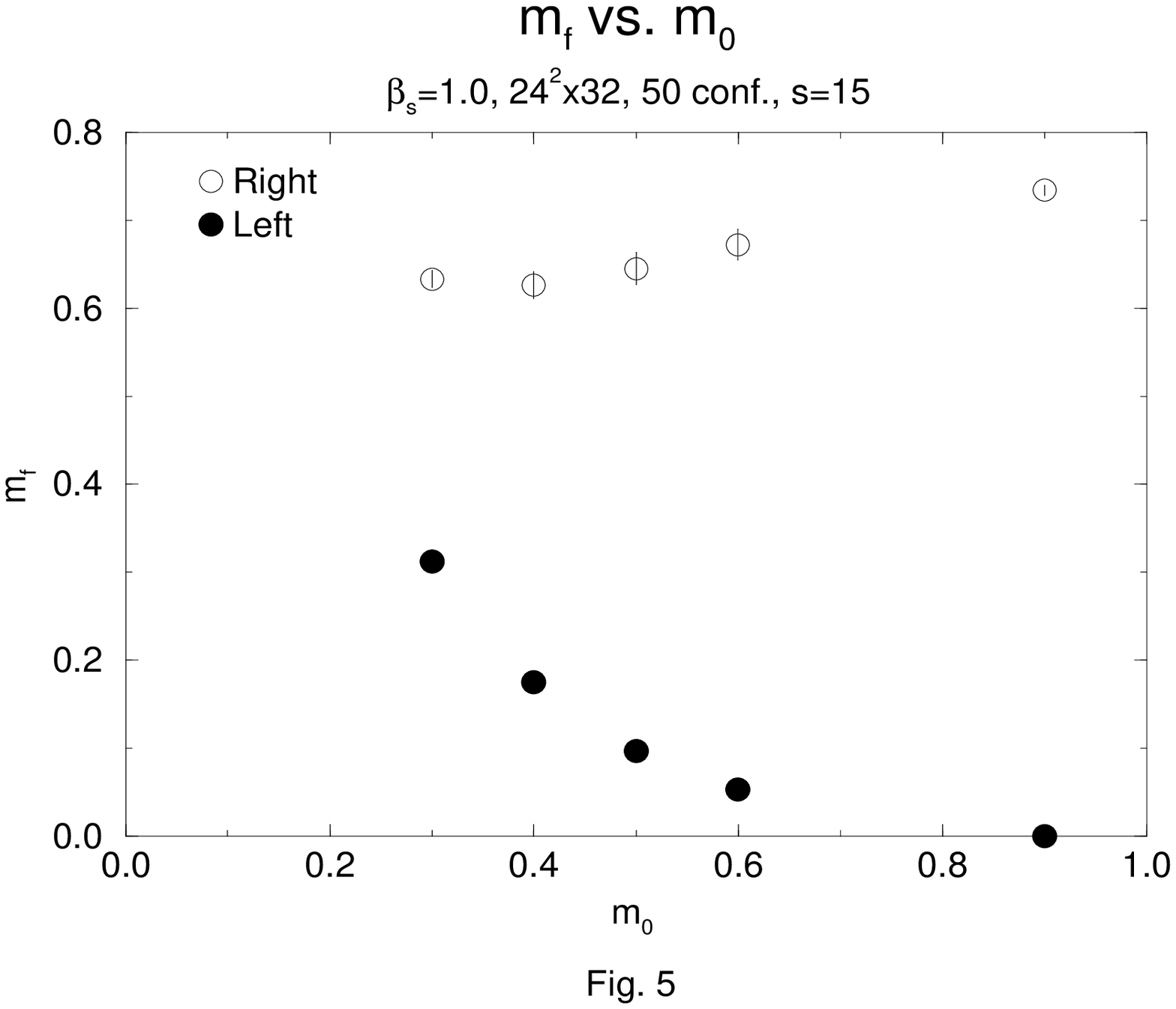}}
\caption{
$m_{f}$ vs. $m_{0}$
at $\beta_{s} = 1.0$ on a $24^{2} \times 32$ lattice,
in the case of putting a source on the anti-domain-wall $s=15$,
for the right-handed fermion(open circles) and the left-handed
fermion(solid circles).}
\label{mfs15}
\end{figure}

On the anti-domain-wall $(s=15)$,
on the other hand,
the mass of left-handed fermion becomes less than 0.1
at $m_{0}$ larger than the critical mass $m_{0}^{c} \sim 0.5$,
as seen in Fig. \ref{mfs15}.
It is noted
that chiralities between the zero modes on the domain-wall and the
anti-domain-wall
are opposite each other.

\begin{figure}
\centerline{\epsfxsize=12cm \epsfbox{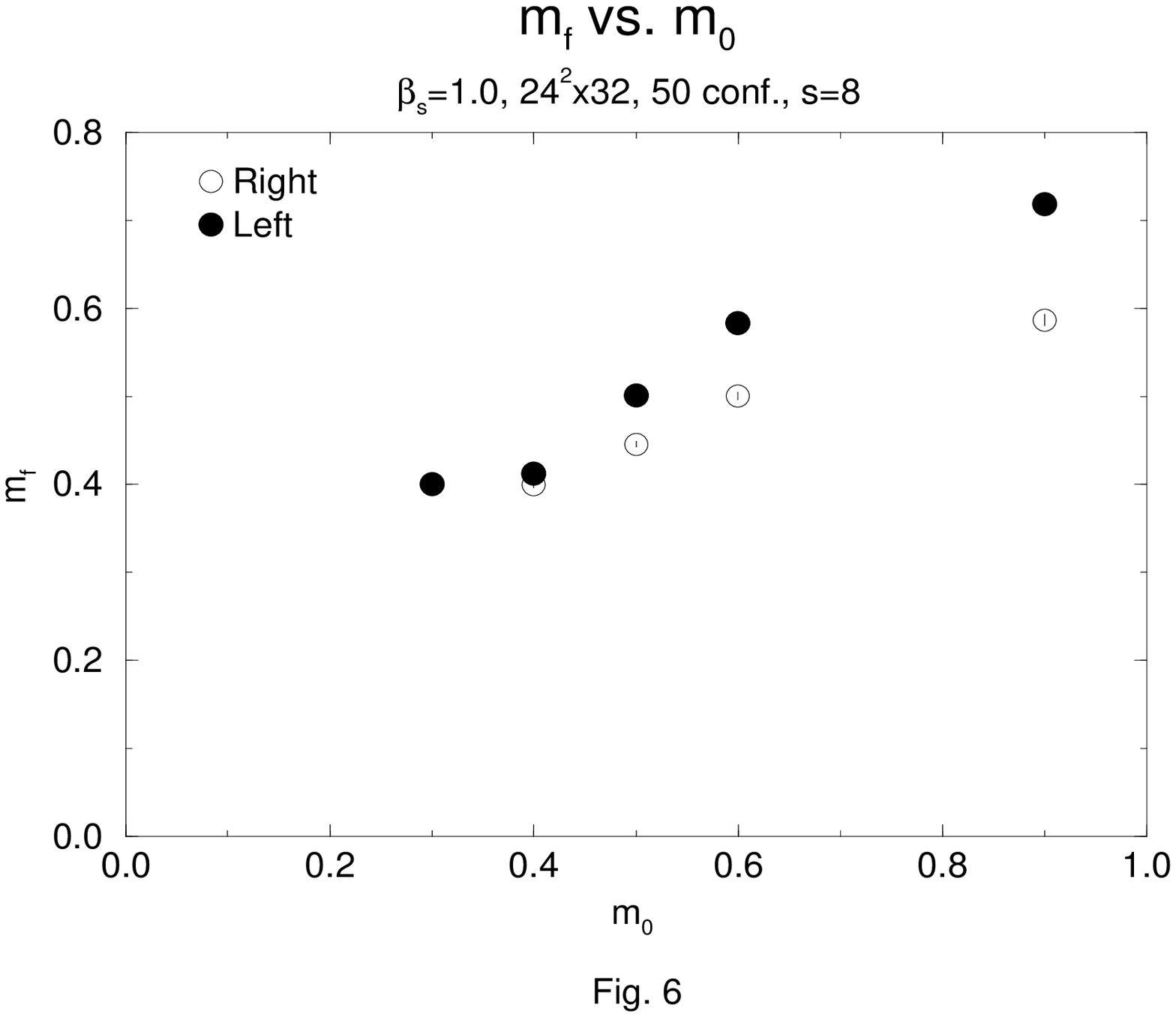}}
\caption{
$m_{f}$ vs. $m_{0}$
at $\beta_{s} = 1.0$ on a $24^{2} \times 32$ lattice,
in the case of putting a source on $s=8$,
for the right-handed fermion(open circles) and the left-handed
fermion(solid circles).}
\label{mfs8}
\end{figure}

Finally Fig. \ref{mfs8} shows that, on $s=8$ ,
the layer in the middle between the domain-wall and the anti-domain-wall ,
both right-handed and left-handed fermions
stay heavy.

\begin{figure}
\centerline{\epsfxsize=12cm \epsfbox{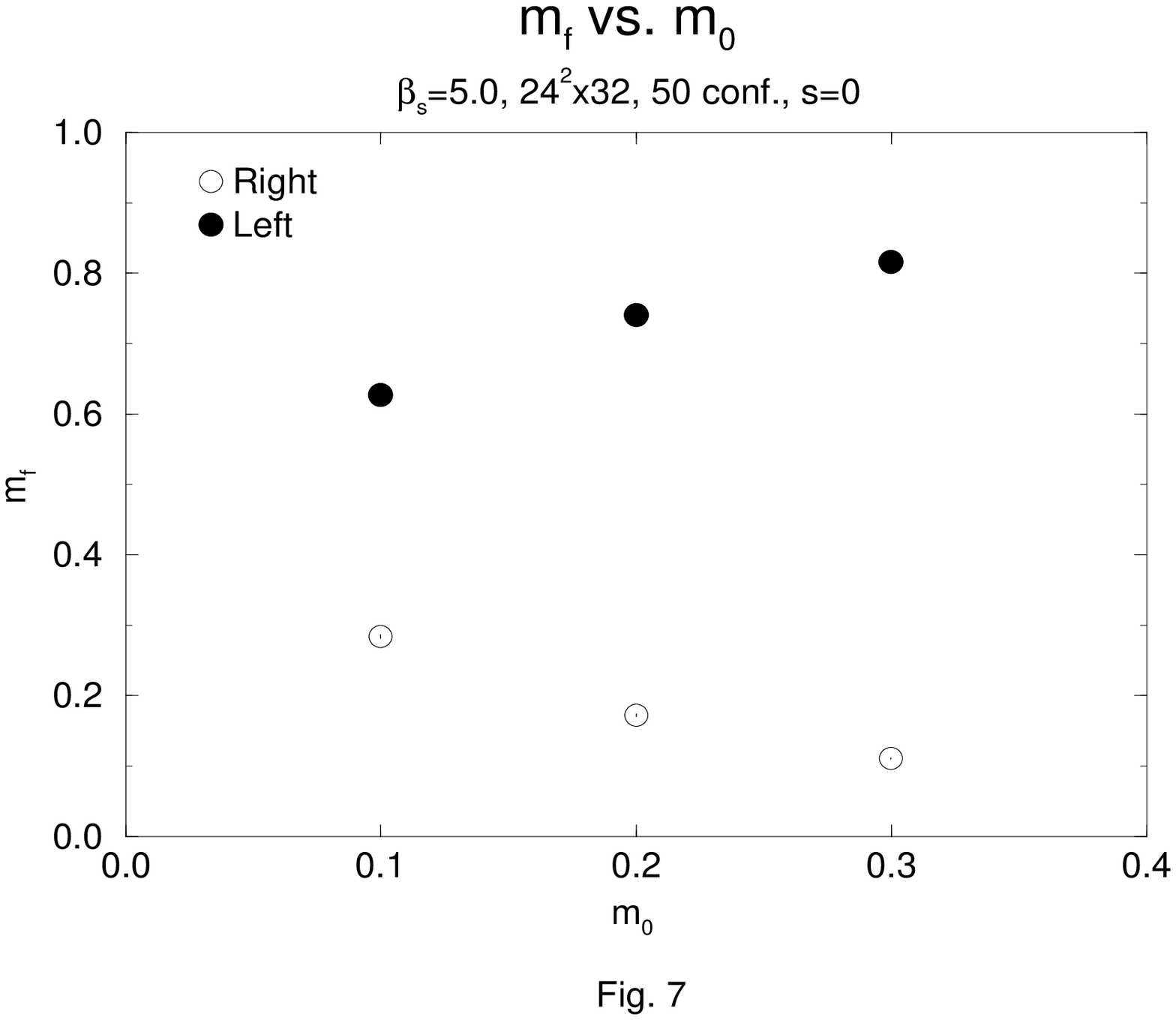}}
\caption{
$m_{f}$ vs. $m_{0}$
at $\beta_{s} = 5.0$ on a $24^{2} \times 32$ lattice,
in the case of putting a source on the domain-wall $s=0$,
for the right-handed fermion(open circles) and the left-handed
fermion(solid circles).}
\label{mfsb5}
\end{figure}

A similar result at $\beta_{s}=5.0$ on $s=0$ is given Fig. \ref{mfsb5}.

{}From these results above,
we conclude
that the domain-wall model with the dynamical gauge field
on the extra dimension ({\it {i.e.}}
the weak coupling limit of the original Kaplan's model)
can create the chiral zero mode on the domain-wall,
at least in the broken phase.
This suggests that the original Kaplan's model
has a great chance to work for the construction of
lattice chiral gauge theories
in the broken phase.

%
%
\subsection{Fermion spectrum in symmetric phase}
\begin{figure}
\centerline{\epsfxsize=12cm \epsfbox{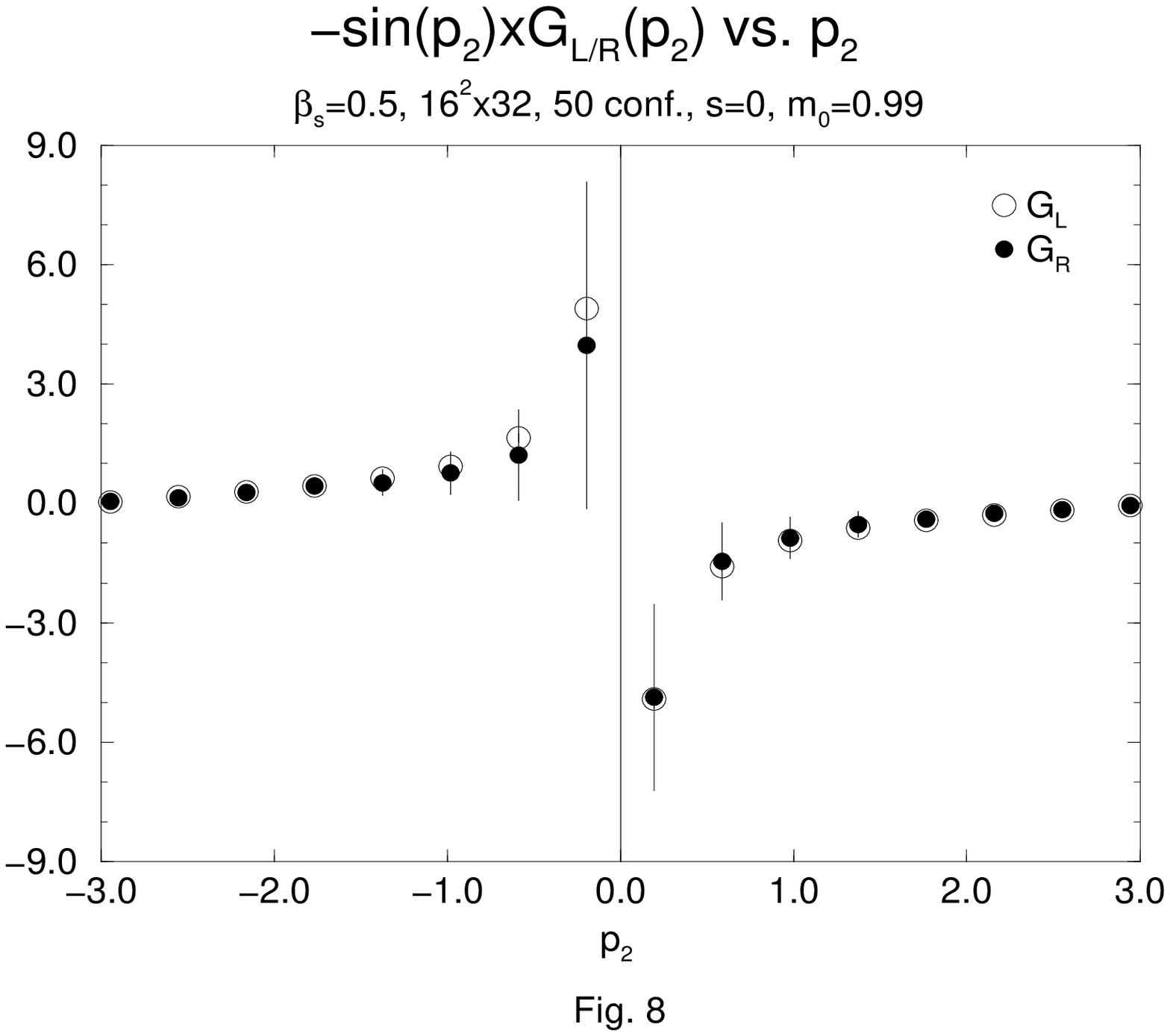}}
\caption{
$ - \sin(p_2) \cdot [G_L]_{0,0}$  (open circles)
and $ - \sin(p_2)\cdot [G_R]_{0,0}$ (solid circles)
as a function of $p_{2}$ with $p_1=0$ at $\beta_{s}=0.5$ on
a $16^2\times 32$ lattice.}
\label{symprop}
\end{figure}

The system is in the symmetric phase at $\beta_{s} = 0.5$.
The fermion propagator is analyzed in the same way as in the broken phase.
However, for example on the $s=0$ layer,
$- \sin (p_{2})\cdot G_{L}$ and $- \sin (p_{2})\cdot G_{R}$
show similar behaviors on a $16^2\times 32$ lattice,
as seen in Fig. \ref{symprop}.
Smaller lattice sizes, stronger the similarity,
which makes analysis more difficult in the symmetric phase.
To see a difference between the right-handed and left-handed fermions,
we have to take larger lattice size such as $L =$ 24, 32.

\begin{figure}
\centerline{\epsfxsize=12cm \epsfbox{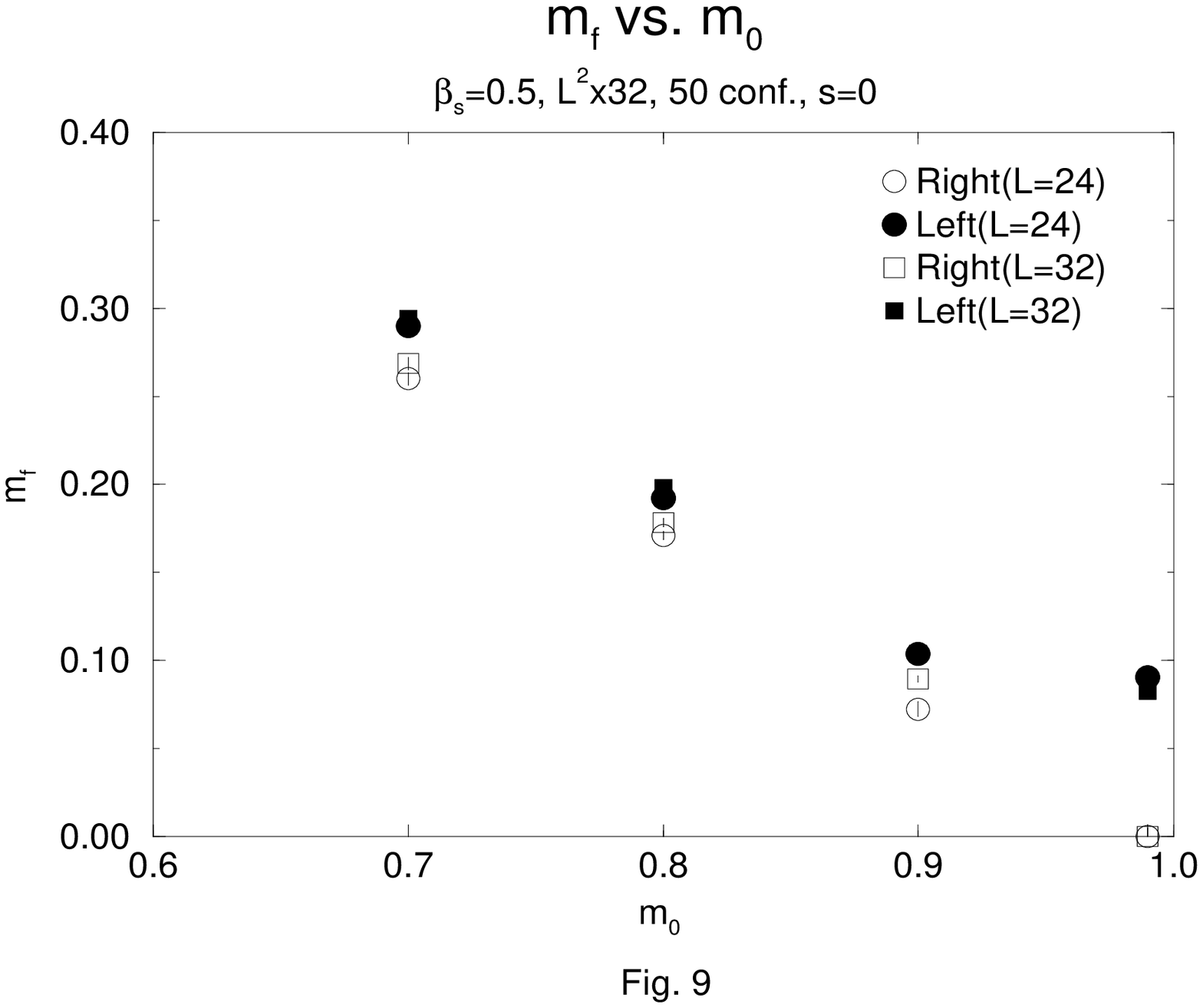}}
\caption{
$m_{f}$ vs. $m_{0}$
at $\beta_{s} = 0.5$ on $L^{2} \times 32$ lattices
with $L=$24(circles) and 32(squares)
in the case of putting a source on the domain-wall $s=0$.
Open symbol stands for the right-handed fermion and
Solid symbol for the left-handed fermion.}
\label{symmfs0}
\end{figure}

In Fig. \ref{symmfs0}, we have plotted mass $m_f$ of both modes
at $s=0$ as a function of $m_0$.
Although a difference of masses
between the right-handed and the left-handed fermions
is very small, about $0.1$ or less at $m_{0}=0.99$,
this difference stays finite as we increase the spatial lattice size $L$
from 24 to 32.
Therefore we conclude
that the right-handed fermion becomes massless at $m_{0}$ larger than
$0.9$,
while the left-handed fermion stays massive at all $m_{0}$,
so that the fermion spectrum on the domain-wall is {\it chiral}.

\begin{figure}
\centerline{\epsfxsize=12cm \epsfbox{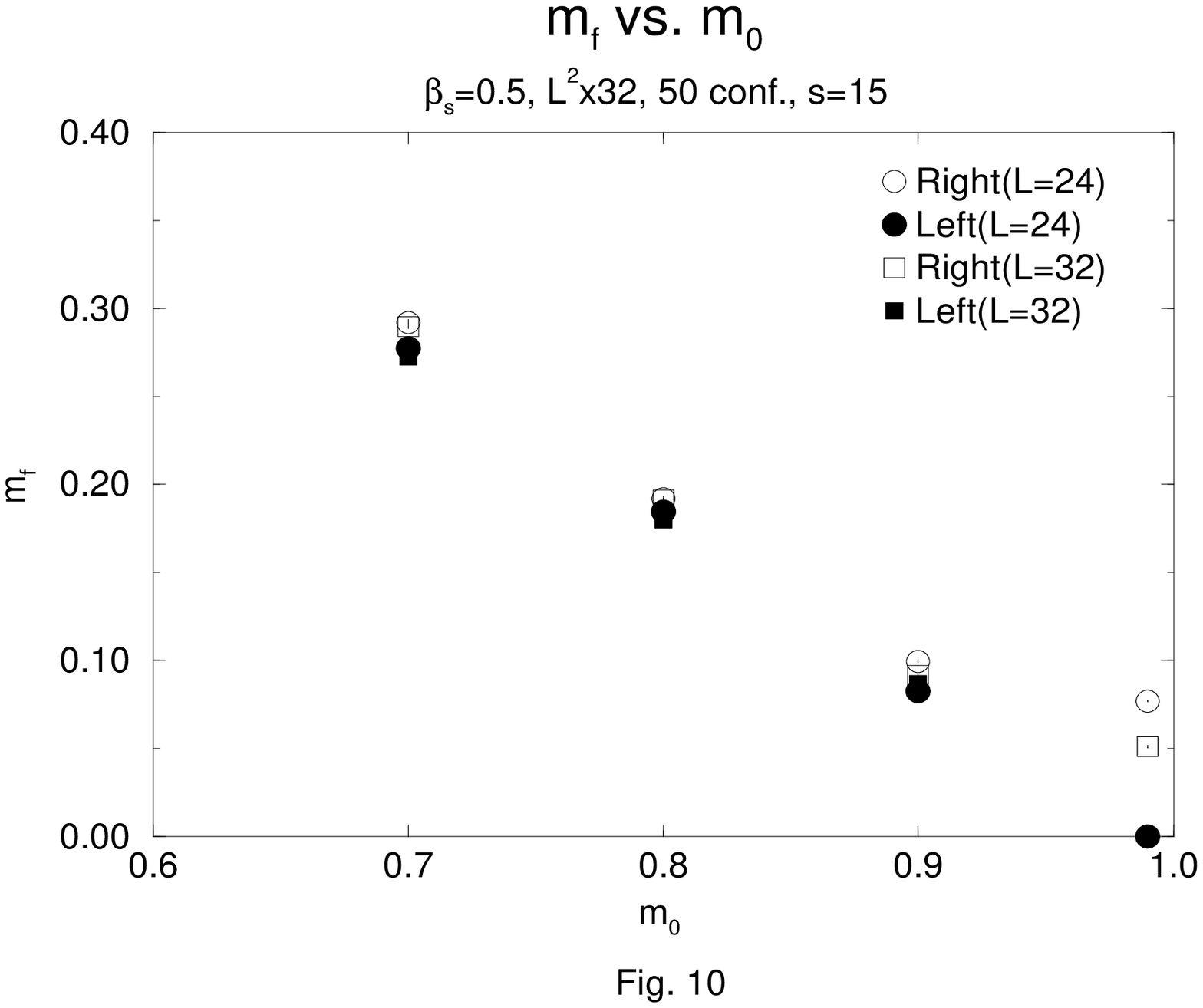}}
\caption{
$m_{f}$ vs. $m_{0}$
at $\beta_{s} = 0.5$ on $L^{2} \times 32$ lattices
with $L=$24(circles) and 32(squares)
in the case of putting a source on the anti-domain-wall $s=15$.
Open symbol stands for the right-handed fermion and
Solid symbol for the left-handed fermion.}
\label{symmfs15}
\end{figure}

In order to see that the difference of mass between the right and the left
is really a signal, not a statistical fluctuation,
we have plotted  $m_f$ vs. $m_0$
in the case of putting a source at the anti-domain-wall $s=15$
in Fig. \ref{symmfs15}.
We observe, at $m_{0}=$0.99,
a massless fermion of the opposite chirality to the $s=0$ zero mode
and a finite difference of masses
between the right and the left, which
stays finite as we increase the spatial lattice size.

\begin{figure}
\centerline{\epsfxsize=12cm \epsfbox{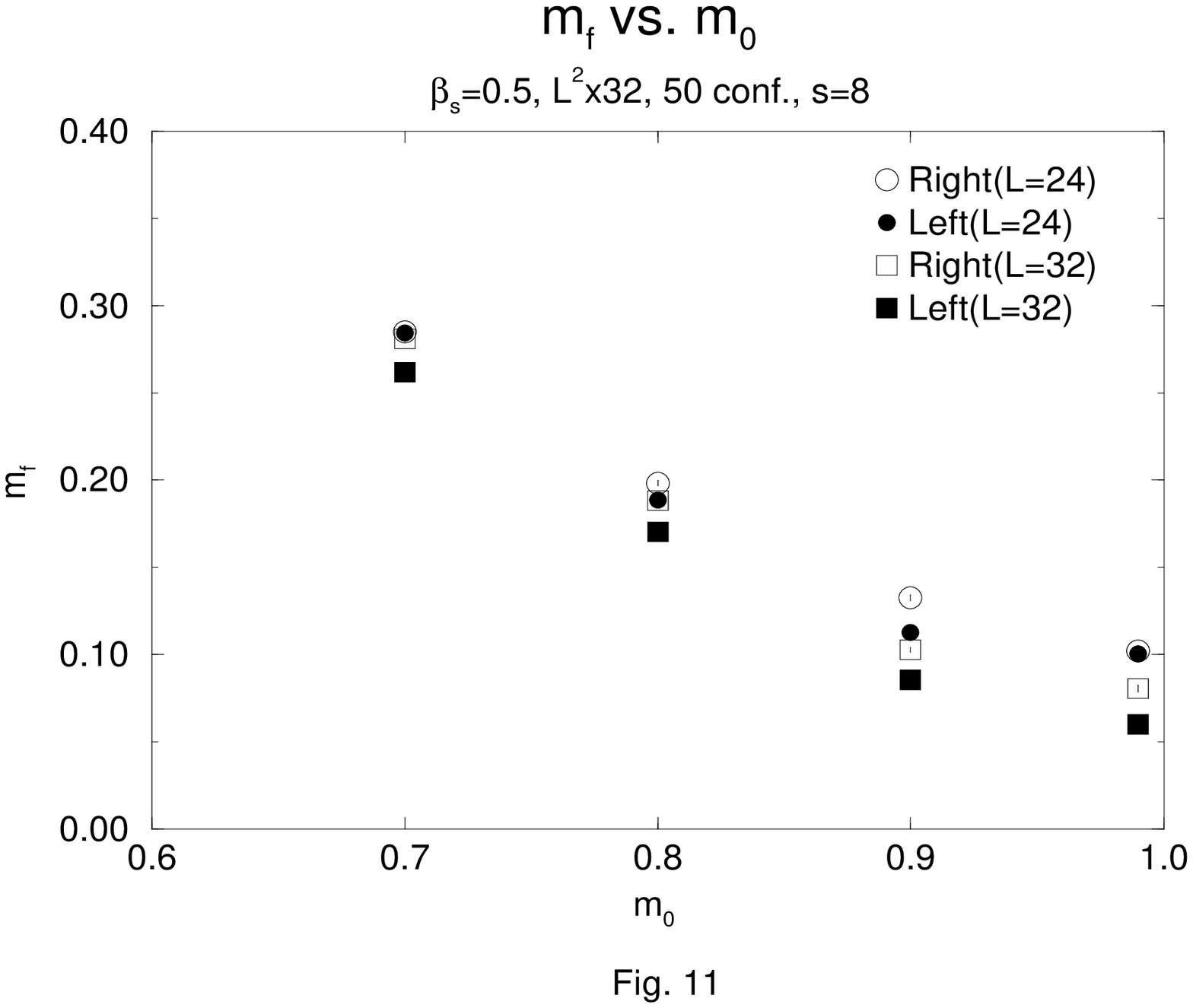}}
\caption{
$m_{f}$ vs. $m_{0}$
at $\beta_{s} = 0.5$ on $L^{2} \times 32$ lattices
with $L=$24(circles) and 32(squares)
in the case of putting a source on $s=8$.
Open symbol stands for the right-handed fermion and
Solid symbol for the left-handed fermion.}
\label{symmfs8}
\end{figure}

Furthermore, in the case of $s=8$,
the right-handed fermion and the left-handed fermion stay massive
at all $m_{0}$, as seen in Fig. \ref{symmfs8}

{}From these results above,
as the same case in the broken phase,
we conclude
that the original Kaplan's model
can create the chiral zero mode on the domain-wall
even in the symmetric phase.

\section{Conclusions and discussions}
Using the quenched approximation,
we have performed the numerical study of the domain-wall model
in (2+1)-dimensions
with the $U(1)$ dynamical gauge field on the extra dimension.
{}From this study we obtain the following results.
There exists the critical value of the domain-wall mass
separating the region with a chiral zero mode
and the region without it, both in the broken and the symmetric phases of
the gauge field.
At the domain-wall mass larger than its critical value
a zero mode with one chirality
exists on the domain-wall
and a zero modes with opposite chirality
on the anti-domain-wall,
and none in the middle between the domain-wall and the anti-domain-wall.

These results strongly suggest
that
it is possible to construct lattice chiral gauge theories
at all $\beta_{s}$ except for $\beta_{s} = 0$
via the domain-wall method, and  continuum limits may possibly be taken
at the critical value of $\beta_s$ where the phase transition takes place.
In $(2+1)$-dimensions, however,
the gauge field in $\beta= \infty$ limit is special
since there is no order parameter and the phase transition is topological.
Thus, to make a definite conclusion on the construction of lattice chiral
gauge theories via the domain-wall method,
we must study realistic $(4+1)$-dimensional model
with $U(1)$ or $SU(N)$ gauge field in $\beta= \infty$ limit.
Such models in $\beta= \infty$ limit have a phase transition
characterized by an order parameter,
a vacuum expectation value of the link variables in the extra dimension.

Moreover, it is interesting and important to find
an appropriate correspondence
between the propagator obtained in the numerical simulation and
the mean field propagator with tuned parameter $z$.
So far it is not clear
what physical quantity is corresponding to $z$.
Since zero mode seems to exist even in the symmetric phase,
the correspondence (\ref{order}) is qualitatively incorrect.
On the other hand, we have found that the correspondence (\ref{energy})
fails to reproduce $m_0^c$ quantitatively.
Since mean-field approximations can not work well in
the lower-dimensions in general,
we must try to answer these questions
studying $(4+1)$-dimensional $U(1)$ or $SU(N)$ models.
%
%
%
\section*{Acknowledgements}
Numerical calculations for the present work have been carried out
at Center for Computational Physics, University of Tsukuba. This work
is supported in part by the Grants-in-Aid of the Ministry of
Education(Nos. 04NP0701, 06740199).
%
%
%
%

%
%
\newpage
%
%

\newpage

\newpage

\newpage

\newpage

\newpage

\newpage

\newpage

\newpage

\newpage

\newpage

\end{document}